\documentclass{aa}
\usepackage{psfig,times}
 
\begin{document} \thesaurus{
 02.14.1; % processes: nucleosynthesis 
 08.19.4; % supernovae: general
 09.09.1: RX J0852.0-4622; % ISM: individual: RX J0852.0-4622 
 09.19.2; % ISM: supernova remnants
 13.07.2; % Gamma-rays: observation
 13.25.4} % X-rays: ISM
\title{Constraints of age, distance and progenitor of the supernova remnant 
RX J0852.0-4622 / GRO J0852-4642}
\subtitle{}
\author{B.Aschenbach, A. F. Iyudin, V. Sch\"onfelder}
 \offprints{B. Aschenbach (bra@mpe.mpg.de)\hfill\break\noindent
}
\institute{Max--Planck--Institut f\"ur extraterrestrische Physik \\
           D--85740 Garching, Germany\\
          }
\date{Received 12 March 1999 /  Accepted 19 July 1999}
\maketitle \markboth{Aschenbach, Iyudin \& Sch\"onfelder: supernova remnant 
 RX J0852.0-4622/GRO J0852-464}{}
\begin{abstract}
The discovery of the nearest young supernova remnant 
RX J0852.0-4622 / GRO J0852-4642
in the Galaxy by ROSAT and COMPTEL has been reported 
recently. Age and distance are determined to $\sim$680 years and $\sim$200 pc by the X-ray diameter 
and the $\gamma$-ray line flux of radioactive $\sp{44}$Ti. Here  we discuss the implications 
of the X-ray spectra and of the fact that 1.8 MeV $\gamma$-ray line emission 
from the decay of $\sp{26}$Al has been measured from the Vela region with a certain fraction 
possibly associated with the new SNR. We estimate an uncertainty of the age of $\pm$ 100 yrs for a fixed 
yield of $\sp{44}$Ti. The highest values of $\sp{44}$Ti yield provided by current supernova explosion models 
give worst case upper limits  of 1100 yrs for the age and of 500 pc for the distance. 
Also the unknown ionization stage of $\sp{44}$Ti adds to the uncertainty of age and distance 
which is at most another 35\% on top. Both the energy balance compiled for the remnant 
and yield predictions for  $\sp{44}$Ti 
and $\sp{26}$Al by supernova models favour a core-collapse event. 
Two point sources have been found in the vicinity of the explosion center, either one of these 
might be the neutron star left by the supernova. If there is a neutron star the 
X-ray count rates of the two point sources provide an upper limit of the blackbody surface 
temperature, which is very unlikely to exceed 3$\times$10$\sp 5$ K. The supernova might 
have been observed some 700 $\pm$ 150 yrs ago, but based on the data of SN 1181, e.g., there is a 
realistic chance 
that it has been missed if the supernova was sub-luminous.        
  \end{abstract}
%
%________________________________________________________________

%
\section {Introduction}
Recently, we have published our discoveries of a previously unknown galactic supernova remnant 
(SNR)  
(Aschenbach 1998, Iyudin et al. 1998). The X-ray image obtained in the ROSAT all-sky survey 
shows a disk-like, partially limb brightened emission region of 2$\sp{\circ}$ in diameter, 
which is the typical appearance of a shell-like SNR 
(cf. Fig.  \ref{picture}). The PSPC X-ray spectra reveal rather high temperatures of  
$>$ 3$\times$10$\sp 7$ K, which indicate that RX J0852.0-4622 is a young object. 
Combining the low age and the 2$\sp{\circ}$ angular extent it is concluded that 
RX J0852.0-4622 is 
relatively close-by. Comparison with historical SNRs limits the age to about 
$\sim$ 1500 yrs and the distance to $<$1 kpc.     
The case of RX J0852.0-4622 being an SNR was clinched by the detection of 
$\gamma$-ray line emission from $\sp{44}$Ti, which is a titanium isotope exclusively 
produced in supernovae. The centre of the $\sp{44}$Ti source, called GRO J0852-4642, is 
off-set from the center of RX J0852.0-4622 by 0.4$\sp{\circ}$, but this is significantly 
less than the angular resolution of the COMPTEL instrument, so that RX J0852.0-4622 
and GRO J0852-4642 are considered to be the same object. 
Using a weighted  mean lifetime
of $\sp{44}$Ti of 90.4 yrs, the angular diameter and adopting a mean 
expansion velocity of 5000 km/s as well as a $\sp{44}$Ti yield of 5$\times$10$\sp{-5}$ 
M$\sb{\odot}$ age and distance are uniquely determined to $\sim$680 yrs and $\sim$200 pc, respectively. 
Therefore, RX J0852.0-4622/GRO J0852-4642 could be the nearest supernova to Earth 
to have occured during recent human history. 

The discovery  of RX J0852.0-4622 and the interpretation as an SNR was made by one of us (BA) 
 in early 1996. During the time which followed it was attempted to associate some fraction 
of the $\sp{26}$Al $\gamma$-ray line emission from the Vela SNR region measured by 
COMPTEL  (Oberlack et al. 
1994, Diehl et al. 1995) with RX J0852.0-4622. The results  have not been 
conclusive basically because of the  unknown distance of RX J0852.0-4622 (Oberlack 1997). 
The discovery of  
$\sp{44}$Ti $\gamma$-ray line emission, however, made it clear that RX J0852.0-4622 is indeed 
a nearby object, so that we could take up again the discussion of the association 
of $\sp{26}$Al emission with RX J0852.0-4622. 
For example the combination of just the $\sp{26}$Al and $\sp{44}$Ti 
data allow to derive  a distance independent estimate of the age of GRO J0852-4642. 
Furthermore, 
if a major fraction of the Vela $\sp{26}$Al mass would be associated with the SNR, a type Ia 
supernova is excluded within the framework of current explosion models.
Under the assumption of adiabatic expansion (Sedov-like) of the SNR we give 
an estimate of the supernova explosion energy E$\sb{0}$ related to the progenitor star 
and the ambient matter density n$\sb{0}$.  
The uncertainties in the determination of age and distance by exploiting the
X-ray spectra are discussed to come up with a time span in which to search for the historical
supernova event. 

In a recent paper Chen \& \ Gehrels (1999) conclude that RX J0852.0-4622
 was created by a core-collapse supernova of a massive star. Their analysis is based 
on the X-ray data and $\gamma$-ray data published published earlier by us (Aschenbach 1998, 
Iyudin et al. 1998). We discuss their approach and conclusions in the
relevant section.

\section {Age and Distance}
Basis for the determination of the age and distance of the new SNR is the law of 
radioactive chain-decay $\sp{44}$Ti~$\rightarrow~\sp{44}$Sc~$\rightarrow~\sp{44}$Ca, for which we 
have
\begin{equation}
 f = {1\over{4\pi d\sp 2}}~{Y\sb{\rm A}\over{m\sb{\rm A}\cdot({\rm{\tau}}\sb{\rm A}-
{\rm{\tau}}\sb{\rm B})}}~
\lbrack exp(-t/{\rm{\tau}}\sb{\rm A}) - exp(-t/{\rm{\tau}}\sb{\rm B})\rbrack
\end{equation}
\noindent with the definitions: {\it f}  photon flux density, {\it d} distance, 
{\it{Y$\sb{\rm A}$}} mass yield of the 
element A, {\it{m$\sb{\rm A}$}} its atomic mass, ${\rm{\tau}}\sb{\rm A}$ its mean life time and 
{\it t} the age. 
For the $\sp{44}$Ti decay chain ${\rm{\tau}}\sb{\rm {Ti}} >> {\rm{\tau}}\sb{\rm {Sc}}$, 
so that we can neglect ${\rm{\tau}}\sb{\rm B}$ and the 
second exponential term of Eq. 1. The data available suggest 
${\rm{\tau}}\sb{\rm{Ti}} \approx$ 90 yrs, which we adopt for the following. This value is 
 also close
 to the  result of (87.7 $\pm$1.7) yrs recently published by  Ahmad et al. (1998).   
{\it f} is the flux of the 1.157 MeV line which has been  measured by Iyudin et al. (1998) to  
(3.8$\pm$0.7)$\cdot$10$\sp{-5}$ 
photons cm$\sp{-2}$ s$\sp{-1}$. 
Apart from the statistical error a systematic error should be added, which is estimated to 
$\pm$ 20\% .
\subsection {$\sp{44}$Ti and X-ray data}
The range of $d$, $t$ and {\it{Y$\sb{\rm A}$}} of Eq. 1 can be constrained by introducing 
the X-results. The angular radius $\theta$ = 1$\sp{\circ}$ is related to $d$ and 
$t$ by $\theta$ = ${v\cdot t}$~/~{\it d}, with $v$ the mean expansion velocity of the SNR. 
By substituting $d$ or $t$ of Eq. 1 by $v$ a quantitative relation between 
$Y\sb{\sp{44}\rm{Ti}}$ and $t$ or 
$Y\sb{\sp{44}\rm{Ti}}$ 
and $d$  can be  derived with $v$ as a parameter. 
An estimate of $v$ can be obtained from the X-ray spectra. 
The analysis of the ROSAT X-ray spectra is affected by the presence of the low
energy emission of the Vela SNR, which is aggravated by the large size of RX J0852.0-4622. 
But an archival ROSAT PSPC pointing observation centred on the
southeastern limb of the Vela SNR with an exposure of 11000 s happens to contain 
 the northern limb section  of RX J0852.0-4622.
The number of counts is sufficient to extract a small section of the limb as well as
a small section of the Vela SNR offset  by just 10 arcmin to create
both uniform source and background spectra.
Limiting the analysis to these small regions, each 10 arcmin $\times$ 10 arcmin in size,
 reduces the impact of any spectral and
spatial non-uniformity across the source and the background.
Fits to the residual northern limb spectrum were performed with a two component, 
optically thin thermal emission equilibrium model (Raymond-Smith model) with 
kT$\sb{\rm l,1}$ =
 0.21$\sb{-0.09}\sp{+0.14}$ keV, kT$\sb{\rm l,2}$ =
 4.7$\sb{-0.7}\sp{+4.5}$ keV 
and an absorption column density of 
N$\sb{\rm H,l,T}$ = 
 $\lbrack$2.3$\sb{-1.5}\sp{+1.5}\rbrack\times$10$\sp{21}$ cm$\sp{-2}$.
 We note that the data can be fit equally well 
($\chi\sp 2\sb{\rm{red}} <$ 1) with a 
straight
power law with index $\alpha$ = --2.6$\sb{-0.4}\sp{+0.3}$ and 
absorption column density 
N$\sb{\rm H,l,\alpha}$ =
$\lbrack$ 1.2$\sb{-1.2}\sp{+7.3}\rbrack\times$10$\sp{20}$ cm$\sp{-2}$.
The spectrum of the rest  of the SNR can be obtained only from the ROSAT all-sky survey data 
and because of the relatively low exposure the spatial region selected for 
analysis needs to be large which increases the uncertainty in assessing the background level 
from the Vela SNR. For the full remnant excluding the bright northern limb an acceptable fit 
with   $\chi\sp 2\sb{\rm{red}} <$ 1 is obtained with 
a two temperature model with  
kT$\sb{\rm r,1}$ =
 0.14$\sb{-0.03}\sp{+0.08}$ keV, kT$\sb{\rm r,2}$ =
 2.5$\sb{-0.7}\sp{+4.5}$ keV and N$\sb{\rm H,r,T}$ =
 $\lbrack$4.0$\sb{-3.5}\sp{+1.5}\rbrack\times$10$\sp{21}$ cm$\sp{-2}$. 

The matter density of the shock-wave heated SNR plasma can be derived from the observed X-ray flux 
$F\sb x$ via the relation 
$F\sb x$ = ${4\over 3}~\theta\sp 3~d~n\sb{\rm e}~n\sb{\rm H}~\Lambda(kT)$.
$\Lambda(kT)$ is the cooling function 
for the best-fit values of $kT$; $n\sb{\rm e}$ is the electron number density and 
$n\sb{\rm H}$ is the number density 
of the un-shocked matter, intitially uniformly distributed in a sphere. Furthermore, 
a factor of four has been used for the density jump at the shock.
The low and high temperature components are associated with densities of 
$n\sb{\rm{H,1}}$ = 0.6$\times$d$\sp{-0.5}\sb 2$ cm$\sp{-3}$ 
and $n\sb{\rm{H,2}}$ = 0.06$\times$d$\sp{-0.5}\sb 2$ cm$\sp{-3}$, respectively, 
with d$\sb 2$ measured in units of 200 pc. Despite the acceptable spectral fit the 
value of $n\sb{\rm{H,1}}$ and the column density are 
 quite uncertain, as the low temperature component could be 
significantly affected by the Vela SNR radiation, which is, however,  not the case 
for $n\sb{\rm{H,2}}$.

\begin{figure}[thb]
  \psfig{figure=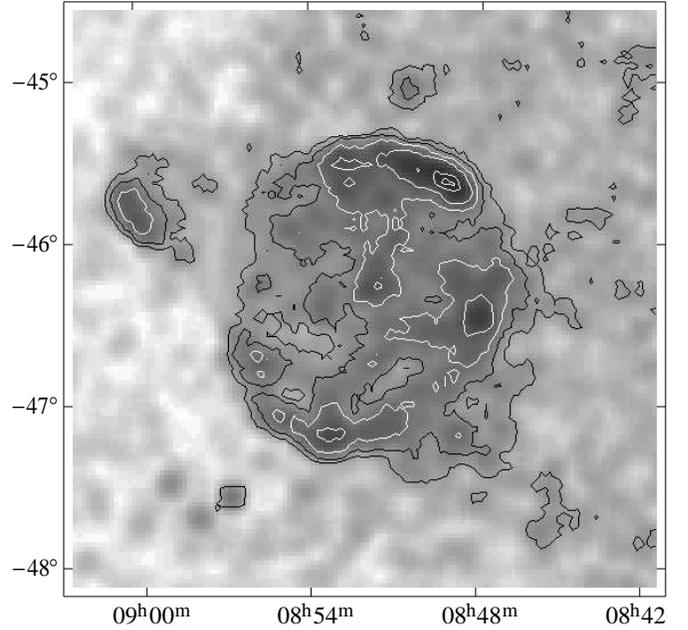,width=8.8truecm,angle=0,%
     bbllx=65pt,bblly=195pt,bburx=520pt,bbury=628pt,clip=}
     \caption[]{Grey scale image of RX J0852.0-4622 for E $>$ 1.3 keV.
      Coordinates are right ascension, declination of epoch 2000.0. 
      Contour levels are (in black) 1.5, 2.3, (in white) 3.5, 5.2, 8.2, 9.2 in units of 
      10$\sp{-4}$ PSPC counts s$\sp{-1}$ arcmin$\sp{-2}$.}
  \label{picture}
\end{figure}

As usual for thermal SNRs two components with different temperatures are needed 
for equilibrium models to fit the observed spectrum. If the plasma 
is far from ionization equilibrium the low temperature 
component appears as an artifact because of the under-ionization. The time-scale 
to reach ionization equilibrium is about 10$\sp{12}$~s~cm$\sp{-3}$/n$\sb e$, with 
n$\sb e$ the electron density of the radiating plasma in units of cm$\sp{-3}$. 
With $t$ = 680 yrs and the densities given above 
1.5$\times$10$\sp 9$~s~cm$\sp{-3}~ <$ n$\sb e\cdot t~<1.5\times$10$\sp{10}$~s~cm$\sp{-3}$, 
which demonstrates that  
 RX J0852.0-4622 departs significantly from ionization equilibrium. Clearly 
the high temperatures observed are closer to the real electron temperature.
But even the high temperatures may underestimate the 
average temperature of the electrons and ions  if the electrons 
are heated mainly by Coulomb collisons with the ions, which occurs on a  
timescale similar to that of reaching ionization equlibrium.
 
The X-ray temperatures which have been produced by shock wave heating can be used 
to estimate the velocity $v\sb s$ of the shock wave: 
$k T = {3\over {16}}~\mu~{\rm{m\sb p}}$$ v\sb s\sp 2$; $\rm{m\sb p}$ is the proton mass and 
$\mu$ is the mean molecular weight, which is 0.6 for a fully ionized plasma of cosmic abundances. 
Again, this relation is for a density jump of a factor of four at the shock.   
Discarding the low temperature components as argued above the X-ray temperatures 
stretch from 1.8 keV $\le$ kT$\sb 2$ $\le$ 9.2 keV including $\pm$1-$\sigma$ errors.
The mean kT$\sb 2$, which is consistent at the same significance level 
with both the radiation from the bulk of the SNR  and its northern limb section, taking the 
thermal option,  is kT$\sb 2$ = 4.4 keV.    
The corresponding best estimates of the minimal and maximal shock velocities 
using the relation above are 1940 km/s, 1240 km/s and 2800 km/s, respectively.
The current shock velocity $v\sb s$ is related to the mean expansion velocity $v$ by the past temporal 
evolution of the SNR. With the limited  
observations available we are forced to rely on  what is known about historical remnants. 
The compilation of Strom (1994) provides both maximal internal shock velocity  
and mean expansion velocity and the ratio $v/v\sb s$ is 
1.5 for the Crab Nebula and Cas A, 2.5 for SN 1006 and 3.5 for the Kepler  and  
Tycho SNRs. For a purely adiabatic expansion in a uniform medium of constant matter density
 (the Sedov description) $v/v\sb s$ = 2.5. As Strom has pointed out the observed maximal internal 
velocities may not be strictly related to $v\sb s$ but they  provide a reasonable estimate. 
More recently measurements of the expansion rate in the X-ray band have become available by 
comparing images obtained with the EINSTEIN and ROSAT observatories or even just the ROSAT images
 taken at different epochs.  
Both Koralesky et al. (1998) and Vink et al. (1998) have found an expansion rate of Cas A of 
0.002\% yr$\sp{-1}$ which corresponds to a factor of $\sim$ 1.55 for the ratio of mean expansion rate 
over  current expansion rate. Hughes (1996) has found a similar value for the Tycho SNR. In each of these 
cases, however, the current expansion velocities, using reasonable distance estimates, are significantly 
larger  than the X-ray spectra and temperatures indicate. Therefore a factor of 1.5 for $v/v\sb s$ 
is a very conservative lower limit to estimate $v$ from X-ray spectra.      

For a worst case estimate we define a velocity range for  
RX J0852.0-4622 applying   factors of 1.5, 2.5, 3.5 to the minimal, best-estimate and 
maximal  $v\sb s$, respectively, which leads to a best-estimate expansion velocity $v\sb{b}$ = 
5000 km/s bracketed by a minimal expansion velocity of $v\sb{min}$ = 2000 km/s and 
a maximal expansion velocity $v\sb{max}$ = 10000 km/s; the values have been rounded off 
slightly. Since we don't know whether the X-ray temperatures are 
associated with either the blast wave heated ambient medium or the progenitor ejecta 
heated by reverse shocks, the expansion velocities derived may even be lower limits. 
Similarly, the values are too low if the electrons have not reached thermal equilibrium 
with the ions.   
For the discussion of the impact of the expansion velocity  
on age and distance we note that the  
 $\sp{44}$Ti line appears to be  broadened (Iyudin et al. 1998), the origin of which 
is unknown. In the most extreme case that the width is exclusively attributed 
to Doppler broadening the associated velocity is (15300$\pm$3700) km/s with an upper 
limit of $v\sb{\gamma}$ = 19000 km/s. We include $v\sb{\gamma}$ for the sake of 
completeness, but we stress that the use of $v\sb{\gamma}$ for constraining the type of
progenitor is overinterpretating the $\gamma$-ray data and it is essentially misleading. 
Nevertheless, we add that Nagataki (1999) quotes ${\sp{44}\rm{Ti}}$ expansion velocities   
$\ge$12000 km/s for sub-Chandrasekhar mass models of SNe Ia. 

\begin{figure}[thb]
  \psfig{figure=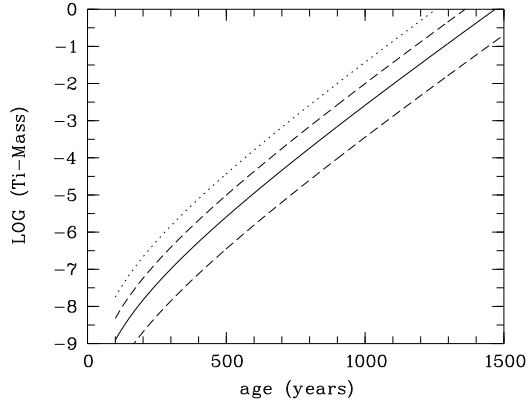,width=6.9truecm,angle=-90,%
     bbllx=45pt,bblly=65pt,bburx=450pt,bbury=585pt,clip=}
     \caption[]{Logarithm of $\sp{44}$Ti yield in solar masses vs. age.
      Lines are for $v\sb{\rm b}$ = 5000 km/s (solid), 
      $v\sb{\rm{min}}$ = 2000 km/s 
      and $v\sb{\rm{max}}$ = 10000 km/s (dashed) and $v\sb{\gamma}$ = 19000 km/s (dotted).}
  \label{age}
\end{figure}

\begin{figure}[thb]
  \psfig{figure=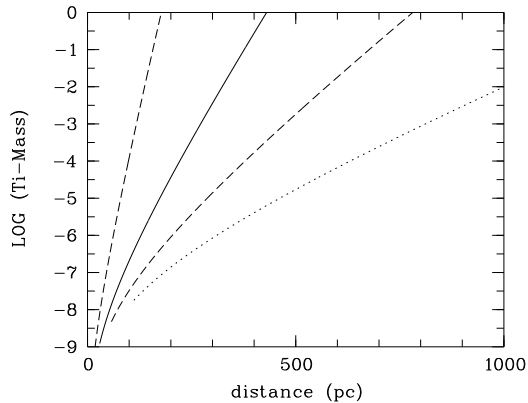,width=6.9truecm,angle=-90,%
     bbllx=45pt,bblly=65pt,bburx=450pt,bbury=585pt,clip=}
     \caption[]{Logarithm of $\sp{44}$Ti yield in solar masses vs. distance.
      Lines are for $v\sb{\rm b}$ = 5000 km/s (solid),
      $v\sb{\rm{min}}$ = 2000 km/s
      and $v\sb{\rm{max}}$ = 10000 km/s (dashed) and $v\sb{\gamma}$ = 19000 km/s (dotted).}
   \label{distance}
\end{figure}

Fig. \ref{age} shows the relation between 
$Y\sb{\sp{44}\rm{Ti}}$ 
and the age $t$ of RX J0852.0-4622 parametrized by $v$. 
Despite the large uncertainty  of $v$, $t$ is determined to within 
$\pm$ 100 yrs for fixed 
$Y\sb{\sp{44}\rm{Ti}}$ 
using $v$ of the X-ray data. 
For 
$Y\sb{\sp{44}\rm{Ti}}$ 
= 5$\times$10$\sp{-5}$
M$\sb{\odot}$ and $v$ = $v\sb b$, $t$ = 680 yrs and $d$ = 200 pc (c.f. Fig. \ref{distance}).
Age and distance are rather insensitive to the exact value of the $\sp{44}$Ti $\gamma$-ray line flux. 
A total error of the flux of $\pm$40\% , which is the sum of the statistical error and the 
systematic error, broadens the range of $t$ by $\pm$40 yrs and that of $d$ by $\pm$ 10 pc.    
If RX J0852.0-4622 is expanding as fast as $v\sb{\gamma}$ indicates the age would be as low 
as $\sim$500 yrs. Model calculations provide a range for 
$Y\sb{\sp{44}\rm{Ti}}$, 
which 
runs for symmetric core-collapse supernovae from 1.4$\times$10$\sp{-5}$ M$\sb{\odot}$ 
to 2.3$\times$10$\sp{-4}$ M$\sb{\odot}$, 
depending on progenitor mass (Woosley \& Weaver 1995, Thielemann et al. 1996). 
Within this range  ($t, d$)  is within (500 yrs, 400 pc) and (950 yrs, 80 pc).  
Nagataki et al. (1998) have pointed out that $Y\sb{\sp{44}\rm{Ti}}$ could be much higher 
in an axisymmetric collapse-driven supernova. For example, 
in  their model with the highest degree of asymmetry  
they obtain $Y\sb{\sp{44}\rm{Ti}}$ 
= 5.1$\times$10$\sp{-4}$ for $Y\sb{\sp{56}\rm{Ni}}$ = 0.07 M$\sb{\odot}$.
This value of $Y\sb{\sp{44}\rm{Ti}}$ would allow an age of up to 1000 yrs.
For type Ia supernovae we find 7.9$\times$10$\sp{-6}$ M$\sb{\odot}$ 
$\le 
Y\sb{\sp{44}\rm{Ti}}$
 $\le$ 4.7$\times$10$\sp{-5}$ M$\sb{\odot}$ for  
carbon deflagration models (Nomoto et al. 1984, Iwamoto et al. 1999), which do not significantly 
differ from the core-collapse SNe, and accordingly the range for ($t, d$) is not affected.  
Values of 
$Y\sb{\sp{44}\rm{Ti}}$ 
as high as 2$\times$10$\sp{-3}$ M$\sb{\odot}$ 
are obtained in He-detonation models (Woosley \& Weaver 1994). This would allow 
($t, d$) to lie between (850 yrs, 500 pc) and (1100 yrs, 100 pc). In summary, for 
any 
$Y\sb{\sp{44}\rm{Ti}}$ given by current models 
 the upper limit of the distance of RX J0852.0-4622 is 500 pc and 1100 years for the age.

Chen \& \ Gehrels (1999) have also used the X-ray temperature to derive age and distance, 
although in a slightly different manner. They use the temperature derived 
from the ROSAT data for the central region (Aschenbach 1998) which might not be 
representative for the current expansion velocity at the rim, why we prefer 
a somewhat higher velocity consistent with the apparently higher temperature observed at 
the limb. For the mean expansion velocity, which should be higher than the current 
expansion velocity by some factor, Chen \& \ Gehrels derive a range of 2000 -- 5000 km/s, 
whereas we propose a range of 2000 km/s to 10000 km/s by comparison with observational data 
obtained for the historical remnants. For given velocity and $Y\sb{\sp{44}\rm{Ti}}$ our results 
agree with those obtained by Chen \& \ Gehrels, but our estimates allow a wider range 
of age and distance.
   
\subsection{$\sp{44}$Ti ionization}
$\sp{44}$Ti decays by electron capture, which means that lifetime depends on ionization stage, 
in particular to what extent the K shell is populated. The $\sp{44}$Ti lifetime of $\approx$ 90 yrs 
is the mean lifetime for two electrons in the K-shell irrespective of the number of 
electrons in the higher shells. For just one electron in the K-shell, the hydrogen-like 
state $\sp{44}${Ti}$\sp{+21}$, the lifetime is expected to be about twice as long, and for the fully ionized 
atom $\sp{44}${Ti}$\sp{+22}$ the lifetime is $>>\rm{\tau}\sb{\rm{Ti}}$.   
Eq. 1 gives the decay rate for the ionic fraction $X(\sp{44}${Ti}$\sp{+22})$ = 0  
(full ionization), $X(\sp{44}${Ti}$\sp{+21})$ = 0 (one electron in the K-shell) and 
$X(\sp{44}${Ti}$\sp{\le +20})$ = 1. 
For $X(\sp{44}${Ti}$\sp{\le +20})$$<$ 1 Eq. 1 is modified by introducing 
the ionic fraction $X(\sp{44}${Ti}$\sp{\le +20})$ with $\tau$  = 90 yrs and $X(\sp{44}${Ti}$\sp{+21})$ 
with $\tau$  = 2 $\times$ 90 yrs; the impact of $X(\sp{44}${Ti}$\sp{+22})$ has been neglected because 
of its comparatively low contribution to $f$. 
The solution for $t$ of Eq. 1 for either the 'ionization' or the 'no-ionization' case is done with 
the same $f$. As before also $d$ and $t$ are not independent of each other but constrained 
by $\theta$ and $v$, which means that not only $t$ but also $d$ is to change for the 'ionization' case 
compared to the 'no-ionization' case. So the comparison is done  with 
the same $v$ but  not with the same $d$. Furthermore $v$ is constrained by  the X-ray spectra and the impact of 
the uncertainty of $v$ on   
$d$ and $t$  has been given in the previous section. 
If $t$ = $t\sb 0$ for $X(\sp{44}\rm{Ti}\sp{\le +20})$ = 1 and   $t$ = $t\sb 1$ for  
$X(\sp{44}\rm{Ti}\sp{\le +20})~\ne$1 Eq. 2 describes the change of the age in terms of 
$q = t\sb 1/t\sb 0$. 

\begin{equation}
q\sp 2 = X(\sp{44}\rm{Ti}\sp{\le +20})~exp[-t\sb 0/\rm{\tau}\sb{\rm{Ti}}\cdot (1-q)]
\end{equation}
$$
        +{1\over 2}\cdot X(\sp{44}\rm{Ti}\sp{+21})~
exp[-t\sb 0/\rm{\tau}\sb{\rm{Ti}}\cdot (1-q/2)]
$$  
 
The ionization stage of $\sp{44}$Ti of GRO J0852-4642 
is not yet known, but a case study is useful to demonstrate  
quantitatively the impact of the ionization on the estimate of $t$ and $d$. 
If $\sp{44}$Ti would have been 
heated to around k$T$ = 4.4 keV like the X-ray emitting plasma, e.g. by a reverse shock 
propagating in the ejecta and if $\sp{44}$Ti is in ionization equilibrium 
the ionic fractions 
can be extracted from literature. Titanium has not been tabulated 
so far but the distributions of the ionic fraction of calcium and iron are available, 
which are taken as case representative examples. Arnaud \& Rothenflug (1985), for instance,
computed $X(\rm{Ca}\sp{\le +18})$ = 0.086, $X(\rm{Ca}\sp{+19})$ = 0.339 and 57.5\% of Ca completely 
ionized for log T = 7.8. Using Eq. 2 $t\sb 0$ = 680 yrs increases to $t\sb 1$ = 930 yrs as 
does $d$ by the same factor of $q$.  
For iron, which has $X(\rm{Fe}\sp{\le +24})$ = 0.686 and $X(\rm{Fe}\sp{+25})$ = 0.269 at log T = 7.8, 
 $t\sb 1$ = 900 yrs, which is very close to the result obtained for Ca,  despite a significantly different 
distribution of the ionic fractions. 
For higher temperatures, e.g. log T = 8.5, $X(\rm{Ca}\sp{\le +18})$ = 0, 
$X(\rm{Ca}\sp{+19})$ = 0.0465 and  95.4\% of Ca is completely
ionized. For this ionic fraction distribution  $q$ = 1 and $t$ and $d$ are unchanged although only 4.65\% 
of the total $Y(\sp{44}\rm{Ti})$ decays radioactively. For even lower values of $X(\sp{44}\rm{Ti}\sp{+21})$, 
i.e. a larger fraction of totally ionized $\sp{44}$Ti, $q~<~1$ or the age becomes even lower.
Using the distribution of the ionic fractions of  iron at log T = 8.5, $q$ = 1.28.          
Clearly, the ionization of Ti has an impact  but of moderate size. 
Values of $t$ and $d$ may be underestimated by some 30\% when the ionization starts 
to affect the K-shell population and they may be even unchanged if only some 10\% or less 
of the Ti has just one electron in the K-shell but is otherwise completely ionized. 

Quite recently Mochizuki et al. (1999) have modelled the heating and ionization of $\sp{44}$Ti 
by the reverse shock in Cas-A, for which they report the possibility of  a currently 
 increased $\sp{44}$Ti activity. 
With respect to RX J0852.0-4622 / GRO J0852-4642 they find that the reverse shock does not heat 
the ejecta to sufficiently high temperatures to ionize $\sp{44}$Ti because of the low ambient 
matter density.  
Future X-ray spectroscopy measurements may answer the question of ionization. But independent  
of the outcome this   
 section shows that even if ionization were significant it does not change the conclusion that
RX J0852.0-4622 / GRO J0852-4642 is a young nearby SNR.      

\subsection{Explosion energy E$\sb 0$}

\begin{figure}[thb]
  \psfig{figure=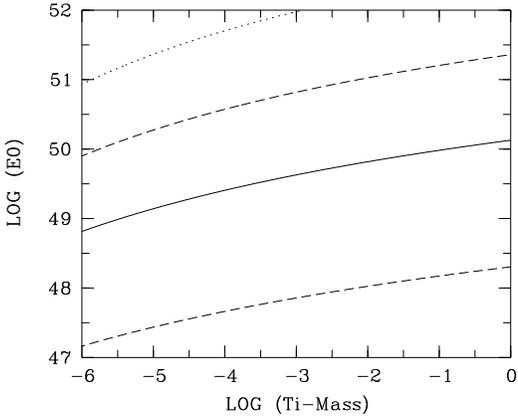,width=6.9truecm,angle=-90,%
     bbllx=45pt,bblly=65pt,bburx=450pt,bbury=570pt,clip=}
     \caption[]{Logarithm of supernova explosion energy E$\sb 0$ in ergs
      vs. logarithm of $\sp{44}$Ti yield in solar masses.
      Lines are for $v\sb{\rm b}$ = 5000 km/s (solid),
      $v\sb{\rm{min}}$ = 2000 km/s
      and $v\sb{\rm{max}}$ = 10000 km/s (dashed) and $v\sb{\gamma}$ = 19000 km/s (dotted).}
   \label{energy}
\end{figure}

The Sedov relation $R\propto~(E\sb 0/\rho\sb 0)\sp{1/5}\cdot~t\sp{2/5}$, which has been 
adopted for  describing the adiabatic expansion of an SNR of radius R in a 
homogenous medium of matter density $\rho\sb 0$, has been used quite often in the past to 
estimate the explosion energy E$\sb 0$ associated with the supernova 
(Winkler \&  Clark 1974, Pfeffermann et al. 1991). 
The limitations of this approach are well known. 
The X-ray spectra provide $k~T$, from which 
 $v$ is derived,  the X-ray flux is proportional to $\rho\sb 0\cdot~d\sp{-0.5}$ via 
$\Lambda (k~T)$ (cf. 
Sect. 2.2) and the X-ray image shows the angular extent $\theta$. 
With the Sedov relation $E\sb 0$ is not yet uniquely determined but can then be expressed 
as a function of a single variable, for instance $t$. Since for RX J0852.0-4622 
$t$ can be  related to 
$Y\sb{\sp{44}\rm{Ti}}$
 via Eq. (1), $E\sb 0$  is a function 
of 
$Y\sb{\sp{44}\rm{Ti}}$. 
 In contrast to the relation $E\sb 0(t)$, $E\sb 0(
Y\sb{\sp{44}\rm{Ti}})$
is constrained because of the limited range of 
$Y\sb{\sp{44}\rm{Ti}}$, at least towards the higher end. 
Fig. \ref{energy} shows $E\sb 0(
Y\sb{\sp{44}\rm{Ti}})$
 for various $v$, because $v$ is not  
uniquely determined by the X-ray spectra.

For the reference values of 
$Y\sb{\sp{44}\rm{Ti}}$
 = 5$\times$10$\sp{-5}$ M$\sb{\odot}$ and 
$v$ = $v\sb b$ = 5000 km/s, $E\sb 0$ = 2.6$\times$10$\sp{49}$, which is a factor 
of about 40 less than the canonical $E\sb 0$ = 10$\sp{51}$ erg.
With $v$ = 5000 km/s this value can not be reached with a realistic 
$Y\sb{\sp{44}\rm{Ti}}$; even a value of $E\sb 0$ = 10$\sp{50}$ erg is hardly consistent with 
reasonable $Y\sb{\sp{44}\rm{Ti}}$ values at $v$ = 5000 km/s. It is interesting to note that 
a similarly low value of $E\sb 0$, i.e. $E\sb 0$ $>$ 4.4$\times$10$\sp{49}$ erg~s$\sp{-1}$ 
have been derived by Willingale et al. (1996) for the SNR of SN 1006,  
 with which RX J0852.0-4622 shares a number of other 
similarities like the X-ray appearance and the ratio of radio to X-ray surface brightness 
(Aschenbach 1998). 
For the reference value of $d$ = 200 pc the total
swept-up mass of RX J0852.0-4622 is less than one solar mass, which means
that the slow-down of the remnant expansion may  not be dominated by
$\rho\sb 0$, so that the applicability of the Sedov relation may be questioned.
The radial evolution  depends then on the details of the explosion rather than just
on $E\sb 0$. For instance, most of the kinetic energy of the SN may be in matter which 
does not radiate in X-rays.

$E\sb 0$ could be raised  by increasing 
$\rho\sb 0$. The slow down could have occurred at  
times $<~t\sb 0$ when regions of higher density might have been passed by the shock wave, e.g. if 
the progenitor star had produced a strong stellar wind. For a mass loss rate of 10$\sp{-5}$ M$\sb{\odot}\cdot$yr$\sp{-1}$
and a wind velocity of 1000 km/s the wind number density would exceed 10$\sp{4}$cm$\sp{-3}$ within a radius 
of about 6$\times$10$\sp{15}$ cm and X-rays would have been emitted from this region. Lower wind velocities like 
those typical of red supergiants would increase the size accordingly.  
The emission region would expand to a measurable size over the 700 yrs but both radiative cooling and adiabatic expansion are likely to 
 have reduced the flux below the detection limit. Nevertheless, we point out that the ROSAT image 
shows weak but enhanced emission from the central 15$\sp{\prime}$ diameter region (c.f. Fig. \ref{picture}).

For higher values of $v$, e.g. for $v$ = $v\sb{max}$ = 10000 km/s and 
$Y\sb{\sp{44}\rm{Ti}}$
 = 5$\times$10$\sp{-5}$M$\sb{\odot}$,   
$E\sb 0$ = 2.9$\times$10$\sp{50}$ erg is relatively close to the canonical E$\sb 0$, 
but the swept-up, X-ray radiating mass is still just 1 M$\sb{\odot}$. Basically, because of the low 
$F\sb x$ and the maximal value of $d$ consistent with 
$Y\sb{\sp{44}\rm{Ti}}$ the swept-up, X-ray radiating mass  never exceeds a few solar masses. 

In summary, E$\sb 0$ is not very sensitive to $Y\sb{\sp{44}\rm{Ti}}$ (c.f. Fig. \ref{energy}) but instead to the 
mean expansion velocity $v$. Taking the full range of $v$ indicated by the X-ray spectra it follows that  
10$\sp{49}$ erg $<$ E$\sb 0$ $<$ 3$\times$10$\sp{50}$ erg for a Sedov-type expansion.

The energy budget made of E$\sb 0$ and the kinetic and thermal energy observed can be used 
to constrain the mass of the progenitor. 
 The total energy E$\sb x$ of the X-ray radiating mass, i.e. the sum of the kinetic energy and the thermal energy, 
amounts to E$\sb x$ = 4$\times$10$\sp{48}$ erg/s$\cdot v\sb{s, 1}\sp 2\cdot d\sb 2\sp{2.5}$ with $v\sb{s, 1}$ 
 in units of 1000 km/s. Since the maximum velocity of the ejecta should not exceed the uniform 
expansion velocity {\it v}, which for the adiabatic case is 2.5$\times v\sb s$, a lower limit of the ejecta  
mass M$\sb{ej}$ is $M\sb{ej}~\ge 100\cdot(\rm E\sb{0, 51}\cdot v\sp{-2} - 6.4\times ~10\sp{-4}~d\sb 2\sp{2.5})$ 
with M$\sb{ej}$ in M$\sb{\odot}$ and E$\sb{0, 51}$ in 10$\sp{51}$ erg. 
For $v$ = $v\sb b$ = 5000 km/s $M\sb{ej}~\ge$ 4  M$\sb{\odot}$, i.e. a massive progenitor is 
required for E$\sb{0, 51}$ = 1, whereas a low mass progenitor with M$\sb{ej}~\ge$ 0.9  M$\sb{\odot}$  
is consistent with the data for $v$ = $v\sb{max}$ = 10000 km/s. A more massive progenitor is required 
if the bulk of the ejecta mass is moving at significantly lower velocities.

Another approach to constrain the progenitor and the supervova type has been taken by Chen \& \ Gehrels 
(1999). They have used the shock wave velocity indicated by the X-ray temperature 
observed in the central region of the SNR and used this as the current expansion  
velocity. By comparison of this velocity with that predicted by SN explosion models 
and their subsequent evolution into an ambient medium of constant matter density, they conclude 
that the likely progenitor of the SNR was a massive star of 15 M$\sb{\odot}$ with a type II
explosion,  solely based on the relatively low value of the current expansion velocity v$\sb{\rm{s}}$ 
inferred from 
the X-ray temperature. Lower mass progenitors like those leading to a SN of type Ia are supposed 
to have significantly higher ejecta velocities 
and according to Chen \& \ Gehrels an ambient matter density $\ge$ 500 cm$\sp{-3}$ is required 
to decelerate the explosion  
wave from initially 11000 km/s to the current value of 1300 km/s, 
using the relation v$\sb{\rm{s}} \propto$ t$\sp{-2/5}$ (Chen \& \ Gehrels, 1999).
If we use the standard Sedov-Taylor relation of v$\sb{\rm{s}} \propto$ t$\sp{-3/5}$ instead, 
i.e. the asymptotic limit of the evolution into a uniform medium of constant matter 
density, a much lower ambient density of 1.4 cm$\sp{-3}$ is sufficient to reduce the ejecta  
speed from 11000 km/s to v$\sb{\rm{s}}$ = 3900 km/s (the upper limit of v$\sb{\rm{s}}$ estimated by Chen \& \ Gehrels) 
in 1000 years for an ejecta mass of one solar mass 
and E$\sb 0$ = 10$\sp{51}$ erg. Although this ambient density still exceeds the observed 
value by a factor of $\sim$30 it is not unreasonable in comparison with other SNRs and  
 ISM densities. Different ejecta mass, explosion energy and non constant density distributions, 
in particular, might reduce the required matter density 
further. In contrast to Chen \& \ Gehrels we are therefore very reluctant to rule out a SNIa 
for RX J0852.0-4622 based on just the X-ray temperature. 
         
\subsection{$\sp{44}$Ti, $\sp{26}$Al and the supernova type}

After the discovery of its X-ray emission in early 1996 it was attempted to 
identify RX J0852.0-4622 as a source contributing to  the 1.8 MeV $\sp{26}$Al $\gamma$-ray line 
emission from the 
Vela region,  
 which 
had been mapped with the COMPTEL instrument (Oberlack et al.
1994, Diehl et al. 1995). Because of its identifaction as an SNR and 
because of its apparently low distance 
RX J0852.0-4622 was considered a good candidate to provide a measurable amount 
of the $\sp{26}$Al $\gamma$-ray line emission. 1.8 MeV $\gamma$-ray lines
are emitted in the radioactive decay of $\sp{26}$Al, which is processed and released 
in supernovae but in other sources as well. The 1.8 Mev Vela source appears to  
be  extended with a significant peak at about 
lII = 267.4$\sp{\circ}$, bII = -0.7$\sp{\circ}$. 
Oberlack (1997) has used the ROSAT X-ray map of the Vela region
 to model the 1.8 Mev $\gamma$-ray map, taking into 
account the full size of the Vela SNR, the Vela SNR explosion fragments (Aschenbach et al. 
1995), RX J0852.0-4622 and other potential sources.
He  found two "COMPTEL point-like" sources which could contribute 
significantly to the $\gamma$-ray peak, which are the Vela SNR fragment 
D/D$\sp{\prime}$ and RX J0852.0-4622.
The peak position and the center 
position of RX J0852.0-4622 agree within the 2-$\sigma$ localization accuracy of COMPTEL, 
and the 1.8 Mev point source flux is  
f$\sb{\rm{Al, m}}$ = (2.2$\pm$0.5)$\cdot$10$\sp{-5}$ photons cm$\sp{-2}$ s$\sp{-1}$ out of the 
total Vela flux of 
f$\sb{\rm{Al}, tot}$ = (2.9$\pm$0.6)$\cdot$10$\sp{-5}$ photons cm$\sp{-2}$ s$\sp{-1}$.
Recently, Diehl et al. (1999) reported a 2-$\sigma$ upper limit of 
2$\cdot$10$\sp{-5}$ photons cm$\sp{-2}$ s$\sp{-1}$
for a contribution of RX J0852.0-4622 to the overall Vela emisson. This result is not really in conflict 
with the result of Oberlack, which we are going to use in the present paper. As we show below most of the 
our conclusions do not depend on the precise value of f$\sb{\rm{Al}}$ anyway.
If f$\sb{\rm{Al, m}}$ is to be attributed to a single SNR with a representative yield 
of $Y\sb{\sp{26}\rm{Al}}$ = 5$\times$10$\sp{-5}$
M$\sb{\odot}$ 
it follows from Eq. (1)
that the distance of the source would be
(160 $\pm$ 20) pc using a mean lifetime of 
${\rm{\tau}}\sb{\rm{Al}}$ = 1.07$\times$10$\sp{6}$
 yrs. Because this excitingly low distance for an SNR was not supported 
by any other measurements at that time and because of other competing 
$\sp{26}$Al sources like the Vela SNR fragment
D/D$\sp{\prime}$ the results were not published.       

But after the discovery of the $\sp{44}\rm{Ti}$ emission, which 
 immediately implies 
a low age because of its
short lifetime and a correspondingly low distance 
because of the X-ray angular diameter, the situation has changed and both the 
$\sp{44}$Ti and the $\sp{26}$Al flux may indeed come from a single supernova now 
visible as the RX J0852.0-4622 SNR. 
Because of the uncertainty of the amount 
of f$\sb{\rm{Al}}$ actually to be attributed to RX J0852.0-4622 we discuss two cases 
in the following chapters: a.) f$\sb{\rm{Al, m}}$ is entirely from 
RX J0852.0-4622; b.) f$\sb{\rm{Al, m}}$ is not entirely 
associated with RX J0852.0-4622 but then  the COMPTEL data provide a firm  upper limit of 
f$\sb{\rm{Al, ul}}$ = 3.5$\times$10$\sp{-5}$ photons cm$\sp{-2}$ s$\sp{-1}$, which is the total 
flux observed for the entire Vela region.

Eq. 1 can be used to compute the age 
$t\sb{\rm{Al,Ti}}$ of RX J0852.0-4622 by using the fluxes of just the two 
radionuclides, making use of $\rm{\tau}\sb{\rm{Ti}} << \rm{\tau}\sb{\rm{Al}}$:
\begin{equation}
t\sb{\rm{Al,Ti}} = \tau\sb{\rm{Ti}}\cdot \rm{ln}(Y\sb{\sp{44}\rm{Ti}}/Y\sb{\sp{26}\rm{Al}}\cdot
26/44\cdot
\tau\sb{\rm{Al}}/\tau\sb{\rm{Ti}}\cdot
f\sb{\rm{Al}}/f\sb{\rm{Ti}})  
\end{equation}
Interestingly, the age determination does not require knowledge of the distance, 
and it depends only on the ratio of the mass yields
of the two elements considered, which might be useful for further searches for young SNRs. 
For f$\sb{\rm{Al}}$ = f$\sb{\rm{Al, m}}$ and $Y\sb{\sp{44}\rm{Ti}}/Y\sb{\sp{26}\rm{Al}}$ = 1,
$t\sb{\rm{Al,Ti}}$ = (750 $\pm$ 25) yrs. This age agrees remarkably well with  the age $t$ = $t\sb 0$ = 
680 yrs which has been derived from the $\sp{44}$Ti data and the X-ray measurements, and it appears 
to   
support the identification of
 RX J0852.0-4622 being the source of both the $\sp{44}$Ti and the $\sp{26}$Al emission. 
Furthermore the value of $t\sb{\rm{Al,Ti}}$ is not very sensitive to the precise value  
of $f\sb{\rm{Al, m}}$; even if only one fifth of $f\sb{\rm{Al, m}}$, e.g., is 
actually associated with RX J0852.0-4622,   
$t\sb{\rm{Al,Ti}}$ is reduced by just 145 yrs.
 
\begin{figure}[thb]
  \psfig{figure=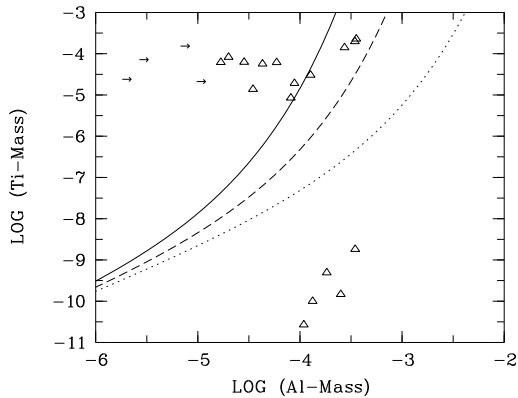,width=6.9truecm,angle=-90,%
     bbllx=45pt,bblly=50pt,bburx=450pt,bbury=580pt,clip=}
     \caption[]{Logarithm of $\sp{44}$Ti yield in solar masses vs. 
      logarithm of $\sp{26}$Al yield in solar masses.
      Lines are for $v$ = 30000 km/s (dotted), 
      $v$ = 10000 km/s (dashed) and $v$ = 5000 km/s (solid). Arrows are for 
      core-collapse model data of 
Thielemann et al. (1996), triangles for model data (S-sequence) of Woosley \& Weaver (1995). }
  \label{limit}
\end{figure}

Some interesting  conclusions can be drawn about the type of the supernova by making use of model produced values 
of ($Y\sb{\sp{44}\rm{Ti}}, Y\sb{\sp{26}\rm{Al}})$.  
The core-collapse models of Woosley \& Weaver (1995) give 0.1 $\le$ $Y\sb{\sp{44}\rm{Ti}}/Y\sb{\sp{26}\rm{Al}}$
$\le$ 4.1 for progenitor masses between 11 M$\sb{\odot}$ and 40 M$\sb{\odot}$ for initial 
solar metallicity, excluding their models with $Y\sb{\sp{44}\rm{Ti}} < 10\sp{-8}$ M$\sb{\odot}$. This leads to 
$t\sb{\rm{Al,Ti}}$ =  (540 -- 880) yrs $\pm$30 yrs. 
Fig. \ref{limit} shows the ($Y\sb{\sp{44}\rm{Ti}}, Y\sb{\sp{26}\rm{Al}}$) -- plane 
of the core-collapse model data (S-sequence of solar metallicity) of Woosley \& Weaver (1995); 
pairs of ($Y\sb{\sp{44}\rm{Ti}}, Y\sb{\sp{26}\rm{Al}}$) with $Y\sb{\sp{26}\rm{Al}}$ 
greater than the values cut by the line of fixed $v$ are not consistent with
f$\sb{\rm{Al, ul}}$.
Fig. \ref{limit} demonstrates that  the models of Woosley \& Weaver (1995) are consistent with relatively low expansion
velocities, most of them
 with $v <$ 5000 km/s, which fits nicely the expansion velocity estimated from the
X-ray temperature.
For increasingly lower metallicity, $Y\sb{\sp{26}\rm{Al}}$ of the Woosley \& Weaver computations
decreases and eventually the data of all the
Z = 0
models are above the $v$ = 5000 km/s cut,  except the models
for which $Y\sb{\sp{44}\rm{Ti}} < $10$\sp{-8}$ M$\sb{\odot}$.
We note that the data of the models S18A, S19A and S25A describing the explosion of the progenitor
with a mass of 18 M$\sb{\odot}$,
19 M$\sb{\odot}$ and 25 M$\sb{\odot}$, respectively,  are closest to the $v$ = 5000 km/s line.
This appears to be in rather good agreement with the conclusion which has been derived
from the energy balance described in
  Sect. 2.3.

The core-collapse models of Thielemann
et al. (1996) for masses between 13 M$\sb{\odot}$ and 25 M$\sb{\odot}$
show similar values of $Y\sb{\sp{44}\rm{Ti}}$ but significantly lower values of $Y\sb{\sp{26}\rm{Al}}$ because
only the yields of the explosively produced elements are given (Thielemann, private communication, 1999). Therefore
$Y\sb{\sp{26}\rm{Al}}$ is to be treated as lower limit, and the applicability to RX J0852.0-4622 remains unanswered at this 
stage. 

Interestingly, Woosley \& Weaver (1995) have also calculated the yields of models with very little 
output of $\sp{56}\rm{Ni}$, from which the supernova power is being drawn 
after a possible 
plateau phase. Models with a small yield of  
$\sp{56}\rm{Ni}$, which may explain the sub-luminous supernovae after the 
early phase, also have low $Y\sb{\sp{44}\rm{Ti}}$ 
but relatively high values of $Y\sb{\sp{26}\rm{Al}}$, 
which is produced predominantly in the upper envelope by ordinary burning.  
The yields predicted by  Woosley \& Weaver
 for solar metallicity are shown in Fig. \ref{limit} as well. 
Clearly, the ratio of $Y\sb{\sp{26}\rm{Al}}/Y\sb{\sp{44}\rm{Ti}}$ is very high and it is too high to be consistent with 
the observations.  
Given the low value of $Y\sb{\sp{44}\rm{Ti}}$,  $Y\sb{\sp{26}\rm{Al}}$ is simply too large.
Such large values have to be checked against f$\sb{\rm{Al, ul}}$.
 f$\sb{\rm{Al, ul}}$
requires a minimal $d$ for a given $Y\sb{\sp{26}\rm{Al}}$, which in turn requires a
maximal $t$ to be consistent with $Y\sb{\sp{44}\rm{Ti}}$.
Minimal $d$ and maximal $t$ define a minimal expansion velocity for the angular diameter not to exceed
$\theta$. Fig. \ref{limit} shows that the Woosley \& Weaver models require very large values of $v$
Such high mean expansion velocities after some 700 yrs are  unlikely and it is evident that these models 
 cannot explain the RX J0852.0-4622 measurements primarily because they are  inconsistent with
 the upper limit of the $\sp{26}$Al flux. Furthermore, the models of Woosley \&  Weaver show a signficant gap for $Y\sb{\sp{44}\rm{Ti}}$, which 
covers the range 3$\times$10$\sp{-8}$ M$\sb{\odot} < Y\sb{\sp{44}\rm{Ti}} <$ 10$\sp{-5}$ M$\sb{\odot}$. 
This gap might be artificial and further model calculations are required to check, whether the fall-back of matter 
towards the center of the explosion chokes the production of the high-Z elements to the extent shown by the 
current explosion models. Models with somewhat lower values of both $Y\sb{\sp{44}\rm{Ti}}$ and $Y\sb{\sp{26}\rm{Al}}$
 would be consistent with the observations, and a lower value of $Y\sb{\sp{44}\rm{Ti}}$ might mean  
 a low value of $\sp{56}\rm{Ni}$ as well, which allows for a sub-luminous supernova 
although the connection between low Y$\sb{\sp{56}\rm{Ni}}$ and low kinetic energy and
 luminosity is not yet well established.

Models for type Ia supernovae predict a much higher ratio of ($Y\sb{\sp{44}\rm{Ti}}/Y\sb{\sp{26}\rm{Al}}$);
Iwamoto et al. (1999) predict 16$< Y\sb{\sp{44}\rm{Ti}}/Y\sb{\sp{26}\rm{Al}} <$ 470  
and the sub-Chandrasekhar models of Woosley \& Weaver have 
280$< Y\sb{\sp{44}\rm{Ti}}/Y\sb{\sp{26}\rm{Al}} <$ 930.   
These values correspond to a relatively large $t$ (Eq. 3) and a low $d$ with 
$f\sb{\rm{Al}}$ = $f\sb{\rm{Al, m}}$, which results in a relatively low value of the mean expansion 
velocity, i.e. 50 km/s $< v <$ 275 km/s for the models of Iwamoto et al. and 180 km/s  $< v < $ 1060 km/s 
for the models of Woosley \& Weaver. 
These values are well below  the lower limit  velocity of $v\sb s$ = 1240 km/s 
and are therefore inconsistent with the X-ray 
temperature measurements.  It appears that 
the type Ia model predictions are in serious conflict with the measurements, thus excluding 
type Ia models from explaining RX J0852.0-4622. But this conclusion hinges on the assumption  that 
$f\sb{\rm{Al}}$ = 
$f\sb{\rm{Al, m}}$ is actually associated with RX J0852.0-4622. If only a minor fraction of 
$<$ 1 \% of $f\sb{\rm{Al, m}}$ is due to RX J0852.0-4622, also type Ia models may be reconsidered.
 For this case $d$ and $t$ are given in Sect. 2.1.  

\section{A compact remnant?}

\begin{figure}[thb]
  \psfig{figure=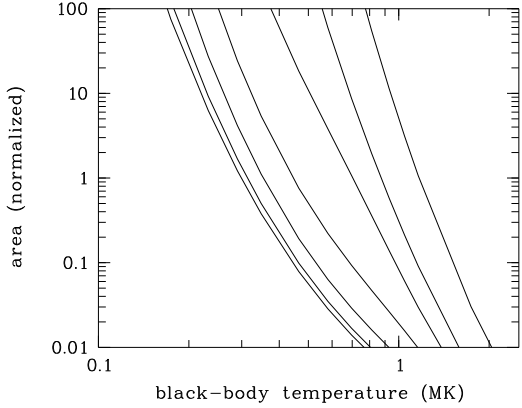,width=6.9truecm,angle=-90,%
     bbllx=45pt,bblly=50pt,bburx=450pt,bbury=565pt,clip=}
     \caption[]{ Black-body surface area normalized to a full size 10 km radius neutron star vs. temperature in units of 10$\sp 6$K. 
Lines are for different interstellar absorption, which is  10$\sp{19}$, 
3$\cdot$10$\sp{19}$, 10$\sp{20}$, 
3$\cdot$10$\sp{20}$, 10$\sp{21}$, 
3$\cdot$10$\sp{21}$, 10$\sp{22}$ in units of cm$\sp{-2}$ from left to right.}
  \label{blackbody}
\end{figure}

In contrast to type Ia supernovae core-collapse supernovae are expected to leave a neutron star 
or a black hole initially close 
to the explosion center. Here we restrict the discussion to a neutron star. 
If borne with a significant kick-velocity the neutron star will travel a distance from the centre  
given by the kick-velocity. Kick-velocities as large as 1000 km/s have been reported for pulsars, and  
for RX J0852.0-4622 with a mean expansion velocity of $v$ = 5000 km/s any putative neutron star should be within a radius 
of about 12$\sp{\prime}$ around the center. The ROSAT all-sky survey data of this area have been searched for  
point sources and two candidate sources have been found. 
Excess emission has been detected  at RA(2000) = 8$\sp{\rm h}$ 52$\sp{\prime}$ 3", DEC(2000) = --46$\sp{\circ}$ 
18$\sp{\prime}$ 36", 
which is off-set from the explosion center by 3.4$\sp{\prime}$. With the nominal value of $v$ and $t$ derived above 
the separation corresponds to 283 km/s for the transverse component 
of the kick-velocity or a proper motion of 0.3" yr$\sp{-1}$. The center of the explosion has been determined by the circle matching 
best the SNR outer boundary. The uncertainty in the center position is estimated to be about $\pm$1.5$\sp{\prime}$. 
This point-like source and the implications concerning a compact remnant have already been reported and discussed (Aschenbach, 1998).
Here we report excess emission from a second point-like source inside the suspected area at 
RA(2000) = 8$\sp{\rm h}$ 51$\sp{\prime}$ 58", DEC(2000) = --46$\sp{\circ}$ 21$\sp{\prime}$ 33". This source has been detected in the 
low energy ROSAT image of the Vela SNR created from the counts which have been recorded in the central 40$\sp{\prime}$ diameter field 
of the PSPC. Compared to the full field of 2$\sp{\circ}$ this procedure improves the spatial resolution considerably and thereby the 
sensitivity of detecting point sources above the diffuse background. The 17 source counts per 40"$\times$40" pixel exceed the 
mean background level of 4.4 counts (40"$\times$40")$\sp{-1}$ by 6-$\sigma$. The source count rate is 0.12 counts/s.          
No spectrum and no information about interstellar absorption is available. But the flux can be used to 
estimate the size of the X-ray emitting area as a function of temperature $T\sb{\rm{bb}}$ for a black-body with the interstellar absorption as 
parameter, which is shown in Fig.\ref{blackbody}. 

If the  source is a black-body radiating neutron star of 10 km radius with emission from the entire surface,  
there is an  upper limit of $T\sb{\rm{bb}}$ = 1.2$\times$10$\sp 6$K 
imposed by the 5-$\sigma$ upper limit of N$\sb{\rm H}$ = 10$\sp{22}$ cm$\sp{-2}$.
But more realistical is a much lower value of N$\sb{\rm H} <$ 10$\sp{20}$ cm$\sp{-2}$, which is typical for the 
southeastern section of the Vela SNR. Then,  Fig.\ref{blackbody} implies 
that  $T\sb{\rm{bb}} \approx$ 3$\times$10$\sp 5$K. If just a fraction 
of the full surface is radiating the temperature may be slightly higher by about a factor of two, for instance, if the area of the 
radiating spot is about 
1\% of the full neutron star surface area. For a full size area radiating neutron star, which is just 700 yrs old, 
$T\sb{\rm{bb}}$ would be surprisingly low. Furthermore, if the supernova left a neutron star somewhere else the black-body surface 
temperature would be even lower because of a lower X-ray count rate, unless the column density
 to the putative neutron star should be even higher. 
 
\section{A historical event?}

A supernova going off at a distance of 200 pc  should have been a spectacular sight for the 
contemporaries in the 13th or 14th century. (If we take into account 
the full band of our age determination also the 12th and the 15th century should  not be 
excluded.) Just how spectacular this event was depends on the absolute 
visual magnitude $M\sb V$. For a bright supernova of type Ia with $M\sb V \le$ --19.0 the burst of light would have   
been as bright as the full moon. For a type II, Ib or Ic which are intrinsically fainter the light output can  
be orders of magnitude less.
 For the sub-luminous or ultra-dim SNe recently discussed (Schaefer 1996, Hatano et al. 1997, Woltjer 1997)  
$M\sb V$ = --13.0 $\pm$ 2. Examples quoted are SN 1181 with $M\sb V$ = --12.68 $\pm$ 1.41 and Cas A 
with $M\sb V =$ --13.03 $\pm$ 2.69. Taking a somwewhat  extreme position with $M\sb V$ = --11.5 the SN associated 
with RX J0852.0-4622 would still have been brighter than Venus. Taking the other extreme position of a type Ia the 
SN is a candidate to be recognized even in daylight.  In any case it should have been seen. 

Records of astronomical events including the epoch of the supernova proposed by us were taken 
by the far-east astronomers of China, Japan and Corea (Clark \&  Stevenson 1977, Ho Peng Yoke 1962).
Their observatories were typically located at a geographical latitude of $\sim$35$\sp{\circ}$ north (Clark \&  Stevenson 1977), 
within $\pm$5$\sp{\circ}$ north or south, so that the 
SN of RX J0852.0-4622 would have risen above the horizon after sun-set by up to 11$\sp{\circ}$ from middle of December to  
end of March.   
If the SN of RX J0852.0-4622 would have exploded say  in late March it would have re-appeared in the northern hemisphere 
during night after more than 250 days with significantly lower brightness. The light curve of the 
sub-luminous SN 1997D (Turatto et al. 1998) provides an  estimate 
of $\Delta m\sb v \sim$ 5 at $\sim$ 250 days after outburst, which appears rather little compared with other SNe 
which show $\Delta m\sb v \sim$ 8 like SN 1994W (Sollerman et al. 1998). With $M\sb V$ = --11.5, $\Delta m\sb v$ =  5 
and a distance modulus of 6.5 the SN of RX J0852.0-4622 then had $V$ = 0 and it is not unlikely to have escaped the attention 
of the medieval far-east  observers. Even with the $M\sb V$ of SN 1181 the chances to miss it would not have been low.
 Furthermore, within this scenario, that only the tail of the light curve had been caught, 
 the SN may have not been noticed as 
a "guest star" because the change was only against a pattern observed more than 200 days before.  
The light curve of SN 1997D also demonstrates that even if the SN went off in the December--March  time frame 
the detection might have been prevented by a rather short peak/plateau period, and sky visibility 
conditions become important. For SN 1997D this period  probably lasted 
for $\le$ 60 days (Turatto et al.), over which the brightness decreased by $\Delta m\sb v \sim$ 3.   

The above exercise demonstrates with realistic data that it is indeed possible that  the SN of RX J0852.0-4622 was not 
bright enough, despite its proximity. This sort of physical explanation requires a sub-luminous SN. 
The peak luminosity and the early lightcurve are determined by the ejecta mass, E$\sb 0$,  
pre-SN radius, the structure of the outer layers, $Y\sb{\sp{56}\rm{Ni}}$ and its distribution. 
As shown by Chugai \& \ Utrobin (1999) in their model for SN 1997D it appears that 
$Y\sb{\sp{56}\rm{Ni}}$ is rather low for this class of sub-luminous SNe, and  
 it remains to be 
seen whether these SNe  can actually produce enough $\sp{44}$Ti
(but c.f. Fig. \ref{age}, \ref{distance} for the minimum amount). We stress that we cannot exclude that the SN was indeed much brighter 
and even a daylight object. Observers located much further south, like the people of the Incas,
the Aztecs or in Middle- and South-Africa, should have had better visibility and their traditions are recommended to be searched for an event  
pointing to a SN. In this context it is interesting to note that the records of the far-east observers as 
published by Ho Peng Yoke (1962) appear to be incomplete. The compilation shows three  gaps, which are 
suspiciously long and statistically inconsistent with the mean rate of entries.  
These periods include the years of   773 -- 814, 1245 -- 1264 and 1277 -- 1293, the latter two of which are 
relevant for  RX J0852.0-4622. 

Finally, we point out that there is a chance that the progenitor star of RX J0852.0-4622 is shown in ancient star charts if it happened to 
be a massive star. Up to now 
just one progenitor star of a supernova has been identified, which is the progenitor of SN 1987A, the B3 Ia blue supergiant 
Sanduleak --69 202 with M$\sb V$ = --6.8 (West et al. 1987). At a distance of 200 pc the apparent unreddened visual magnitude 
would have been  $V$ = --0.3, which would have made the star the brightest star in the Vela constellation located between 
$\gamma$ Vel and $\lambda$ Vel. A Wolf-Rayet type progenitor star would have been less bright with $V \approx$ +2.5, but still comparable 
with the other bright stars in Vela. This opens up an interesting explanation for the apparent absence of a historical record, which admittedly 
is a speculation. If the progenitor star had been so bright the supernova might not have been noted down as a "guest star". 
The existing star would just have become brighter. And if the short peak of the outburst had been missed of whatsoever reason  
and only the tail of the supernova lightcurve has been observed the change of brightness might not have been spectacular, and 
the star would have disappeared slowly over a couple of years.
 
\section{Conclusions}

An estimate of the age $t$ and distance $d$ of the supernova remnant 
RX J0852.0-4622 / GRO J0852-4642 can be obtained by combining the 
ROSAT X-ray and COMPTEL $\gamma$-ray data. Assuming a $^{44}$Ti 
yield of the supenova of 5 $\times$ 10$^{-5}$ M$_{\odot}$ and an 
expansion velocity of 5000 km/sec $t$ = 680 yrs and $d$ = 200 pc are  
obtained. Actually, the expansion velocity is constrained by the 
X-ray data to lie in the range of 2000 km/sec $<$ v $<$ 10 000 km/sec 
yielding an uncertainty of the age of $\pm$ 100 years 
for a fixed $^{44}$Ti yield. For the highest $^{44}$Ti yield given by current 
supernova models, a firm upper limit of the distance is 500 pc and 1100 
years for the age.

The determination of the age depends to some extent  on the ionization 
state of $^{44}$Ti because $^{44}$Ti decays by electron capture.  
The values quoted above have been  obtained under the assumption 
that the K-shell is fully populated. If the K-shell 
contains only one electron, the $^{44}$Ti mean lifetime is estimated to increase by a factor of two. 
But the age of the SNR will not increase by the same factor because 
of the angular diameter, and therefore distance  constraint. Adopting the same mean expansion 
velocity $t$ and $d$ can change by about 35\% at most and for a very strong ionization $t$ and $d$ 
may be even lower than the "nominal" estimate. Future X-ray spectroscopy measurements are needed 
to search for Ti X-ray emission lines to determine the ionization state of $^{44}$Ti and further 
constrain $t$ and $d$.

The X-ray surface brightness of RX J0852.0-4622 is rather low and implies a rather low 
matter density of the shock wave heated plasma if the radiation is thermal. 
A formal analysis of the X-ray data in terms of a Sedov-type evolution of the SNR using 
the standard conversion of X-ray temperature in shock velocity turns out a rather 
low value of a few times 10$\sp{49}$ erg for the explosion energy $E\sb 0$, which can be raised 
only significantly if the mean expansion velocity $v$ would exceed 10000 km/s. But if $v$ is closer 
to 5000 km/s as the X-ray data indicate then the bulk of $E\sb 0$ resides still in kinetic energy of the 
ejecta, not radiating in X-rays, which means that any reverse 
shock has not yet penetrated deep into the ejecta, and that the
 titanium is not highly ionized. In this case a lower limit for the mass of the 
progenitor star of 25 M$\sb{\odot}$ is estimated from the energy balance. 
   
There is evidence 
for $^{26}$Al 1.809 MeV line emission from RX J0852.0-4622, 
which has been
measured by COMPTEL towards the Vela region. Admittedly this has  still to be confirmed. But  
if a non-neglible part of this 1.809 MeV line flux 
is coming from RX J0852.0-4622 
a similar age of 600-750 yrs for the SNR 
is obtained for similar yields of $^{26}$Al and $^{44}$Ti. Since $t$ depends only on the 
logarithm of the yields and the fluxes $t$ will not change significantly even for large 
changes of the $^{26}$Al flux. It is more a matter of whether or not there is $^{26}$Al emission.
If the $^{26}$Al 
line flux is about what is indicated by the COMPTEL data the existing type Ia supernova 
models can be ruled out for the progenitor explosion because they predict 
a ratio of the  $^{26}$Al and $^{44}$Ti yield which is by far too large.  They could  be reconsidered only 
if less than 1\% of the  $^{26}$Al line flux from Vela in total is associated with RX J0852.0-4622.
Explosion models of core-collapse supernovae ( Woosley \& Weaver, 1995) are in general in agreement with 
the observations, i.e. the measurements of  
$v$, $^{44}$Ti line flux and $^{26}$Al line flux, judging from their prediction of the yields of $^{44}$Ti
 and  $^{26}$Al. Their models with $Y\sb{\sp{44}\rm{Ti}} <$ 10$\sp{-8}$ M$\sb{\odot}$ and 
$Y\sb{\sp{26}\rm{Al}} >$ 10$\sp{-4}$ M$\sb{\odot}$ 
 can definitely be excluded because the yield of $^{26}$Al predicted is 
inconsistent with even the upper limit of the $^{26}$Al line flux measured for the Vela region in total.
In summary, both the energy balance and the yield predictions of the currently available explosion models
point towards a core-collapse event.  

Within the vicinity of the explosion center two point-like X-ray emission regions have been found, 
either of which could be the manifestation of a neutron star. Spectra are not available, but if 
the radiation is assumed to be black-body emisson, the X-ray flux indicates a surface temperature 
of $\approx$3$\times$10$\sp 5$ K, which is surprisingly low for 
a 700 years old neutron star. If neither one of these two sources is a neutron star and the neutron 
star hides somewhere else with an even lower X-ray count rate or if only a fraction 
of the radiation observed from the point-like objects is due to thermal radiation the surface temperature       
of the neutron star could be even lower.

In principle, the  supernova could have been seen from the far-east astronomers of China, 
Corea or Japan or from geographical latitudes further south.
The supernova could have been very bright and then there are records expected to exist, which 
should be searched for. If the supernova would have been of the sub-luminous class with a brightness 
as low as that of SN 1181 and a short peak-plateau duration  
it could have been missed. There is some chance that the progenitor star was sufficiently bright and could 
have been seen by the naked eye. Then the stellar pattern of Vela was different in ancient times and 
the astronomers who monitored the sky might not have noted the supernova as a "guest star" or a "new" 
star, because the star was existing and just brightening and eventually fading away.

\begin{acknowledgements}
We would like to thank U. Oberlack and R. Diehl for fruitful discussions on the COMPTEL 
$\sp{26}$Al-data.
\end{acknowledgements}

\vfill\eject 

\begin{thebibliography}{}

\bibitem{}
Ahmad I., Bonino G., Cini Castagnoli G., et al., 1998, Phys. Rev. Lett. 80, 2550 
\bibitem{}
Arnaud M., Rothenflug R., 1985, A\&AS 60, 425
\bibitem{}
Aschenbach B., 1998, Nature 396, 141
\bibitem{}
Aschenbach B., Egger R., Tr\"umper J., 1995, Nature 373, 587  
\bibitem{}
Chen W., Gehrels N., 1999, ApJ 514, L103
\bibitem{}
Chugai N.N., Utrobin V.P., 1999, astro-ph/9906190 10 Jun 1999
\bibitem{}
Clark D.H., Stephenson F.R., 1977, "The Historical 
Supernovae", Pergamon, Oxford 
\bibitem{}
Diehl R., Bennet K., Bloemen H., et al., 1995, A\&A 298, L25
\bibitem{}
Diehl R., Oberlack U., Pl\"uschke S., et al., 1999, Proc. 3rd INTEGRAL Workshop, 
Taormina 1998, eds. G. Palumbo, H. Bazzano, C. Winkler, Astrophys. Lett.\& Comm., 1999, in press
\bibitem{}
Hatano K., Fisher A., Branch D., 1997, MNRAS 290, 360
\bibitem{}
Ho Peng Yoke, 1962, Vistas in Astron. 5, 127
\bibitem{}
Hughes J., 1996, BAAS 189, 4608
\bibitem{}
Iwamoto K., Brachwitz F., Nomoto K., et al., 1999, ApJ subm. 
\bibitem{}
Iyudin A.F., Sch\"onfelder V., Bennett K., et al., 1998, Nature 396, 142
\bibitem{}
Koralesky B., Rudnick L., Gotthelf E.V., Keohane J.W., 1998, ApJ 505, L27
\bibitem{}
Mochizuki Y., Takahashi K., Janka H.-Th., Hillebrandt W., Diehl R., 1999, A\&A 346, 831 
\bibitem{}
Nagataki S., 1999, ApJ 511, 341
\bibitem{}
Nagataki S., Hashimoto M.-A., Sato K., Yamada S., Mochizuki Y., 1998, ApJ 492, L45
\bibitem{}
Nomoto K., Thielemann F.-K., Yokoi K., 1984, ApJ 286, 644 
%\bibitem{}
%Norman, E.B., et al., 1998, Phys. Rev. C 57, 2010 
\bibitem{}
Oberlack U., 1997, PhD Thesis, \"Uber die Natur der galaktischen $\sp{26}$Al-Quellen --
Untersuchung des 1.8 MeV Himmels mit COMPTEL, Technische Universit\"at M\"unchen
\bibitem{}
%Oberlack, U., Diehl, R., Montmerle, T., Prantzos, N., von Ballmoos, P., 
% et al., 1994, ApJS 92, 433
Oberlack U., Diehl R., Montmerle T., et al., 1994, ApJS 92, 433
\bibitem{}
Pfeffermann E., Aschenbach B., Predehl P., 1991, A\&A 246, L28
\bibitem{}
Schaefer B.E, 1996, ApJ 464, 404
\bibitem{} 
Sollerman J., Cumming R.J., Lundqvist P., 1998, ApJ 493, 933
\bibitem{}
Strom R.G., 1994, A\&A 288, L1
\bibitem{}
Thielemann F.-K., Nomoto K., Hashimoto M.-A., 1996, ApJ 460, 408
\bibitem{}
Turatto M., Mazzali P.A.,Young T.R., et al., 1998, ApJ 498, L129
\bibitem{}
Vink J., Bloemen H., Kaastra J.S., Bleeker J.A.M., 1998, A\&A 339, 201
\bibitem{}
West R.M., Lauberts A., J\o rgensen H.E., Schuster H.-E., 1987, A\&A 177, L1
\bibitem{}
Willingale R., West R.G., Pye J.P., Stewart G.C., 1996, MNRAS 278, 749
\bibitem{}
Winkler Jr., P.F., Clark G.W., 1974, ApJ 191, L67
\bibitem{}
Woltjer L., 1997, A\&A 328, L29
\bibitem{}
Woosley S.E., Weaver T.A., 1994, ApJ 423, 371
\bibitem{}
Woosley S.E., Weaver T.A., 1995, ApJS 101, 181
\end{thebibliography}
\end{document}